\documentclass[11pt,tightenlines]{revtex4-1}
\newcommand{\Go}{{G\={o}}}

\newcommand{\kT}{k_B T}

\newcommand{\pr}{p_r}
\newcommand{\pR}{p_R}
\newcommand{\PR}{P_R}

\newcommand{\br}{{\bf r}}
\newcommand{\bp}{{\bf p}}

\newcommand{\bF}{{\bf F}}

\newcommand{\bR}{{\bf R}}
\newcommand{\bP}{{\bf P}}
\newcommand{\bV}{{\bf V}}

\newcommand{\intW}{{\displaystyle \int}\!{\rm d}}

\newcommand{\etal}{{\it{et al.}}}

\usepackage{graphicx}
\usepackage{amssymb}
\usepackage{amsbsy}
\usepackage{amsmath}
\usepackage{mathtools}

\usepackage{epstopdf}
\usepackage[normalem]{ulem}
\usepackage{verbatim}

\begin{document}
\author{Joseph F. Rudzinski}
\email{rudzinski@mpip-mainz.mpg.de}
\affiliation{Max Planck Institute for Polymer Research, Mainz 55128, Germany}

\title{
Recent progress towards chemically-specific coarse-grained simulation models with consistent dynamical properties
}

\begin{abstract}
Coarse-grained (CG) models can provide computationally efficient and conceptually simple characterizations of soft matter systems.
While generic models probe the underlying physics governing an entire family of free-energy landscapes, bottom-up CG models are systematically constructed from a higher-resolution model to retain a high level of chemical specificity.
The removal of degrees of freedom from the system modifies the relationship between the relative time scales of distinct dynamical processes through both a loss of friction and a ``smoothing'' of the free-energy landscape.
While these effects typically result in faster dynamics, decreasing the computational expense of the model, they also obscure the connection to the true dynamics of the system.
The lack of consistent dynamics is a serious limitation for CG models, which not only prevents quantitatively accurate predictions of dynamical observables but can also lead to qualitatively incorrect descriptions of the characteristic dynamical processes.
With many methods available for optimizing the structural and thermodynamic properties of chemically-specific CG models, recent years have seen a stark increase in investigations addressing the accurate description of dynamical properties generated from CG simulations.
In this review, we present an overview of these efforts, ranging from bottom-up parametrizations of generalized Langevin equations to refinements of the CG force field based on a Markov state modeling framework.
We aim to make connections between seemingly disparate approaches, while laying out some of the major challenges as well as potential directions for future efforts.
\end{abstract}
\maketitle

\section{Introduction}

Particle-based, low-resolution molecular simulation models, which represent groups of atoms with a single coarse-grained (CG) site, have been instrumental in shaping our perspective of soft matter systems.
From generic bead-spring polymer models~\cite{Doi:1986} to experimentally-informed \Go-type protein models~\cite{Go:1975,Clementi:2008}, 
representing a system with reduced resolution not only enables access to larger length and longer time scales, but also---and arguably more importantly---can help to elucidate the essential microscopic driving forces of emergent phenomena.
The coarse-graining procedure is rigorously grounded by statistical mechanics, which can be leveraged to develop systematic routines for parametrizing the potential energy function governing interactions at the CG level of resolution as well as the corresponding equations of motion.
As early as the 1930's, both Onsager and Kirkwood discussed the concept of the many-body potential of mean force (MB-PMF)---a state-point-dependent free-energy function which acts as the proper CG interaction potential for reproducing all structural and thermodynamic properties of the underlying system at the CG level of resolution~\cite{Onsager:1933,Kirkwood:1935ys}.
A whole range of ``bottom-up'' techniques have been introduced to approximate this high-dimensional function using simulations of a higher-resolution model, normally with an atomically-detailed and classical representation~\cite{Noid:2013uq}.
These methods typically employ standard molecular mechanics potentials for this purpose, which in most cases correspond to a highly deficient basis set for approximating the MB-PMF.
The reader is directed to~\cite{Peter:2009hr,Riniker:2012qf,Noid:2013uq} for instructive reviews on this topic.

A few decades following the work of Onsager and Kirkwood, Mori and Zwanzig independently developed a framework for mapping the dynamics of a microscopic system onto an arbitrary set of observables that are functions of the phase space variables of the underlying system~\cite{MORI:1965bx,Zwanzig:1961}.
The crux of this theory is a projection operator formalism which results in ``projected dynamics'' that evolve according to a generalized Langevin equation.
In the context of particle-based CG models, this results in the time evolution of CG sites that is not only determined by the MB-PMF, but also by frictional and stochastic forces which account for the missing degrees of freedom.
The latter forces are related through a fluctuation-dissipation theorem and depend in general not only on the coordinates and momenta of the CG sites but also on time.
Similar to methods for approximating the MB-PMF, the Mori-Zwanzig (MZ) framework can be employed to parametrize the CG equations of motion, but only after significantly simplifying the form of the frictional and stochastic forces~\cite{Hijon:2010}.
Despite the well-known result from Mori and Zwanzig, most bottom-up CG models are simulated according to a simple Langevin equation, i.e., with a single, uniform friction coefficient and the corresponding Gaussian distributed white noise for the stochastic force.
This implementation assumes that the frictional forces are Markovian---not explicitly dependent on time---and independent of the coordinates and momenta of the CG sites.
If this rather harsh assumption is valid, then the friction coefficient need not be explicitly calculated and, instead, any CG dynamical property may be interpreted via a uniform time-rescaling factor which relates the sampled CG time scales with the true underlying time scales of the system.
Remarkably, such a rescaling has been demonstrated in certain regimes for both configurational polymer dynamics and transport phenomena in liquids~\cite{Rosenfeld:1999,Padding:2011}.
 
\begin{figure}[htbp]
  \begin{center}
    \includegraphics[width=\linewidth]{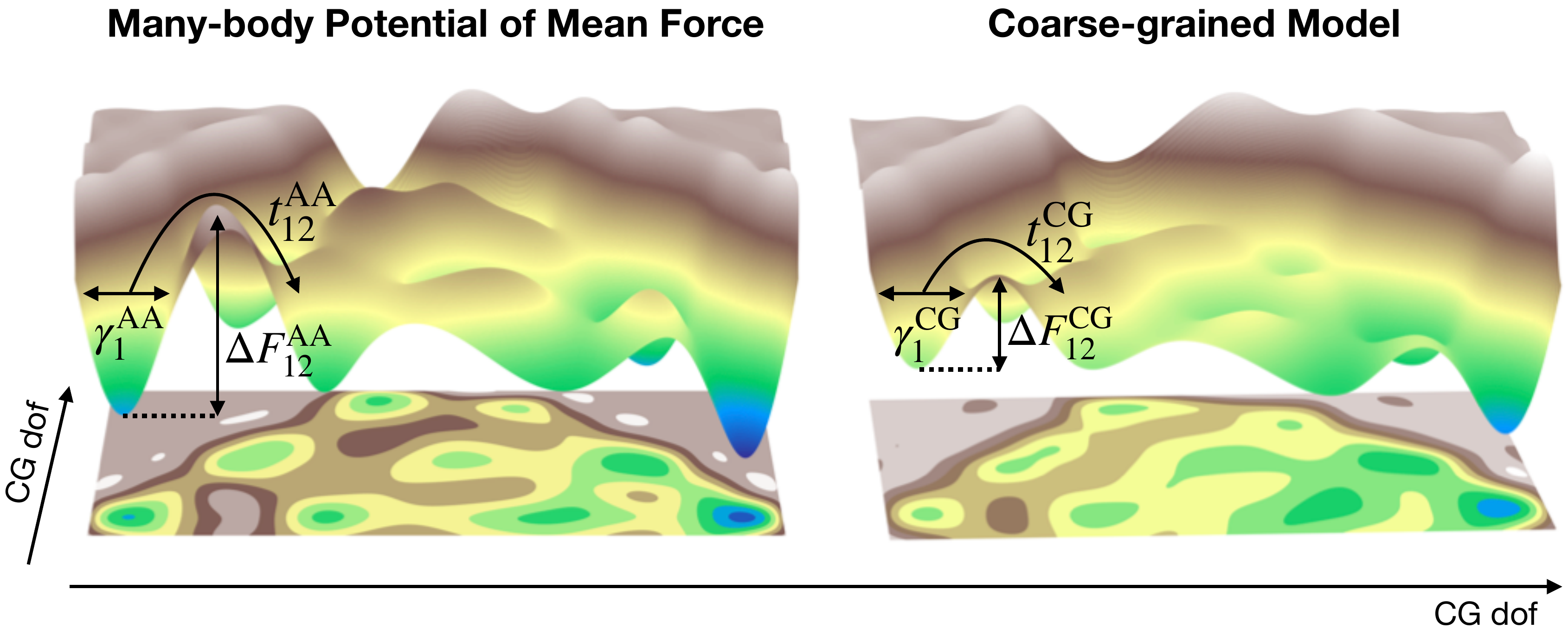}
    \caption{
Illustration of the challenge of CG dynamics.
Free-energy surfaces generated by the MB-PMF (left) and an approximate CG model (right) at the CG level of resolution (i.e., plotted along CG degrees of freedom (dof)).
CG models lose connection with the true dynamics for two distinct reasons: (i) The friction, $\gamma_1$, within a free-energy basin is different from the AA model, even if the MB-PMF is employed, due to the missing degrees of freedom and (ii) the approximate CG model effectively smooths over the MB-PMF, changing the relative barrier heights, $\Delta F_{12}$, between a pair of basins in a non-trivial way.
As a result, not only is the time scale associated with hopping between pairs of basins, e.g., $t_{12}$, not quantitatively reproduced, but the ratio of time scales with respect to two distinct transitions is changed.
Furthermore, these errors propagate into the description of longer time scale processes, severely limiting the predictive capabilities of CG models.
\vspace{-4mm}
}
    \label{fig:FEL}
  \end{center}
\end{figure}

More generally, the time rescaling is not uniform and instead is a complicated function of the dynamical process of interest, as dictated by the nature of the MZ frictional force.
As a consequence, not only is the direct connection to the true dynamical time scales of the system lost, but the distribution of pathways sampled by a CG model may be qualitatively incorrect.
To better understand this concept, it is useful to consider the evolution of the system over the free-energy landscape determined by the CG potential and corresponding equations of motion (Figure~\ref{fig:FEL}).
Even if we were able to perfectly reproduce the true underlying landscape (at the CG level of resolution), i.e., simulate the MB-PMF, a nontrivial MZ frictional force may be necessary to account for the reduced friction resulting from the removal of degrees of freedom.
As already mentioned above, in practice the landscape generated by a CG model is almost always a drastic approximation to the true landscape, since we typically employ a highly deficient basis set of interactions.
This approximation is often interpreted as performing a smoothing of the MB-PMF, since CG models typically employ softer interaction potentials with respect to their all-atom (AA) counterparts.
Smoothing generally results in nontrivial adjustments in the relative barrier heights along the landscape and almost certainly complicates the form of the MZ frictional force.
Thus, both missing frictional forces as well as incorrect relative barrier heights may result in a qualitatively incorrect distribution of pathways sampled by a CG model.

Clearly, the challenge of accurately extracting dynamical properties from CG models is severely limiting for their predictive capabilities.
At the same time, despite the generally daunting nature of the problem as laid out by the MZ framework, CG models have been extensively and successfully employed to gain insight into dynamical processes in soft matter.
This is especially true for more generic CG models, which are often explicitly constructed using physical intuition about the driving forces for the phenomena of interest.
This ``built-in physics'' restricts certain features of the free-energy landscape and, as a result, also the corresponding dominant pathways between states.
For chemically-specific CG models, the situation is complicated by incorporating finer details into the model which, if not performed carefully, may lead to a misleading description of the characteristic dynamical processes.
In this case, researchers have approached the problem by (i) explicitly correcting the dynamics following the MZ framework, (ii) employing empirical rescaling relations, valid for particular classes of systems, to match specific transport properties, or (iii) refining the CG interaction potential in an attempt to represent the hierarchy of long-time scale processes (i.e., kinetic properties) through the reproduction of the dominant free-energy barriers.

In this review, we discuss recent work towards characterizing and improving the dynamical properties generated by CG simulation models, while attempting to build a conceptual connection between seemingly disparate approaches.
We focus on relatively high-resolution CG representations which aim to retain a high level of chemical specificity, thereby complicating the description of both static and dynamical properties of the system.
At the same time, we will occasionally look to representative examples of coarser or more generic models for intuition and inspiration.
The review is organized as follows: Section~2 discusses bottom-up approaches following the MZ formalism, starting with a brief summary of the necessary theoretical foundations; Section~3 presents time-rescaling approaches, with a focus on polymers and liquids; Section~4 covers the free-energy-landscape perspective, including the use of structural-kinetic-thermodynamic relationships and Markov state models to assist in the interpretation and refinement of CG kinetic properties; Section~5 briefly highlights a few representative applications which provide significant challenges for the methods discussed in the preceding sections; Finally, Section~6 provides a brief overarching discussion and outlook for future work.

\section{Bottom-up approaches following the Mori-Zwanzig (MZ) formalism}
\label{Sec-MZ}
In this section, we will review recent work that utilizes the MZ approach to address discrepancies in CG dynamics.
In contrast to the original application of MZ to model the evolution of macroscopic observables~\cite{Zwanzig:1961,MORI:1965bx,Zwanzig:1964,Robertson:1966,Kawasaki:1973,Koide:2008}, e.g., scattering functions~\cite{LiPi:2014}, we focus here on modeling the time evolution of CG sites that represent groups of atoms.

\subsection{Preliminaries}

Consider an atomic system with coordinates $\br = \{ \br_1, \dots, \br_n\}$ and momenta $\bp = \{ \bp_1, \dots, \bp_n\}$ interacting according to a classical potential energy function $u(\br)$ in the canonical ensemble.
The atomistic Hamiltonian is:
\begin{equation}
h(\br,\bp) = \sum_{i=1}^{n} \frac{1}{2m_i} \bp_i^2 + u(\br) \,\, ,
\end{equation}
where $m_i$ represents the mass of atom $i$.
The probability of finding the system in configuration $\br$ is given by: 
\begin{equation}
\pr(\br) = z(\br)^{-1} \exp(-\beta u(\br)) \,\, ,
\end{equation}
where $z(\br) = \intW \br \, \exp(-\beta u(\br))$ is the configuration integral and $\beta = 1 / \kT$ is the inverse temperature.
Let us define transformations from atomic coordinates and momenta to a lower resolution representation: $\bR_I = M_{\bR I}(\br)$ and $\bP_I = M_{\bP I}(\bp)$ for each CG site $I = 1,...,N$.
The mapping functions $\{M_{\bR I}\}$ and $\{M_{\bP I}\}$ are typically chosen as linear functions.
We assume that the CG system also lies within the canonical ensemble with probability of configuration $\bR$ given by: 
\begin{equation}
\PR(\bR) = Z(\bR)^{-1} \exp(-\beta U(\bR)) \,\, ,
\end{equation}
where $Z(\bR) = \intW \bR \, \exp(-\beta U(\bR))$ is the configuration integral and $U(\bR)$ is a classical potential governing interactions at the CG level of resolution.
One natural consistency criterion for the CG model is that the configurational probability distribution should match the distribution generated by the high-resolution model after being mapped to the CG representation~\cite{Noid:2008a}:
\begin{equation}
\PR(\bR) = \pR(\bR) \equiv \intW \br \, \pr(\br) \, \delta(M_{\bR}(\br) - \bR) \,\, ,
\label{eq:cons}
\end{equation}
where $\delta(M_{\bR}(\br) - \bR) \equiv \prod_{I=1}^N \delta(M_{\bR I}(\br) - \bR_I)$.
$\pR(\bR)$ is the probability that an atomic configuration is mapped to a particular CG configuration.
By solving Equation~\ref{eq:cons} for the CG potential, we recover a special potential energy function, $U^0(\bR) \propto -\kT \ln \pR(\bR)$, which is known as the many-body potential of mean force (MB-PMF).
An equivalent consistency criterion can be imposed for the distribution of CG momenta, $\bP$, which is automatically satisfied for certain forms of the mapping functions $\{M_{\bP I}\}$~\cite{Noid:2008a}.

The dynamical evolution of the high-resolution model is determined by Hamilton's equations of motion.
Equivalently, the phase-space distribution, $\rho(\br,\bp)$, evolves according to the Liouville equation:
\begin{equation}
\frac{\partial \rho}{\partial t} + i\mathcal{L}\rho = 0 \,\, ,
\end{equation}
where $\mathcal{L} = \sum_{i=1}^{n} \left ( \partial / \partial \br_i + \partial / \partial \bp_i \right )$ is the Liouville operator.
In the MZ approach, $\mathcal{L}$ is decomposed into two parts using orthogonal operators $P$ and $Q$: $i\mathcal{L} = Pi\mathcal{L} + Qi\mathcal{L}$, where the sum of $P$ and $Q$ is equal to the identity operator.
$P$ projects out the dynamics of the $N$ CG degrees of freedom, which are generally considered to correspond to slowly relaxing degrees of freedom.
Then, the ``missing'' or ``removed'' degrees of freedom (i.e., the degrees of freedom projected out by $Q$) are meant to relax quickly and, thus, can be represented implicitly in the resulting CG model.
One can then derive an expression for the equations of motion of CG sites, which evolve according to a generalized Langevin equation (GLE)~\cite{Kinjo:2007,Hijon:2010,Izvekov:2014,Deichmann:2018}:

\begin{equation}
\frac{{\rm d}\bP}{{\rm d}t}(t) = - \underbracket{\vphantom{\int_{0}^{t} {\rm d}t' \, \Gamma(\bR, \bP, t-t') \bV(t')} \frac{{\rm d}}{{\rm d}\bR} U^0(\bR(t))}_{\text{\large conservative}} - \underbracket{\strut \int_{0}^{t} {\rm d}t' \, \Gamma(\bR, \bP, t-t') \bV(t')}_{\text{\large frictional}} + \underbracket{\vphantom{\int_{0}^{t} {\rm d}t' \, \Gamma(\bR, \bP, t-t') \bV(t')} \delta F^{\rm Q}(t)}_{\text{\large random}} .
\label{eq-GLE}
\end{equation}
The chosen projection operator, $P$, determines the form of the dissipative forces~\cite{Izvekov:2013}---i.e., the frictional and random forces.

In most applications, a Markovian assumption is made to remove the explicit time dependence from the friction kernel: $\Gamma = \Gamma(\bR,\bP)$.
Hydrodynamic-like interactions between CG sites are incorporated by retaining $\Gamma$'s dependence on $\bP$, in contrast to traditional hydrodynamic approaches~\cite{Duenweg:2007} that explicitly represent sites that mimic the effective behavior of the missing degrees of freedom, e.g., solvent.
Most often, the friction kernel is assumed to depend only on the CG coordinates, $\Gamma = \Gamma(\bR)$, and can be further simplified by assuming a particular functional form.
The most widely used approximation is to assume that the friction is pairwise additive and depends only on the distance between two particles.
In this case, Equation~\ref{eq-GLE} simplifies to a form which corresponds to the dissipative particle dynamics (DPD) approach~\cite{Espanol:2017}: 
\begin{equation}
\frac{{\rm d}\bP_I}{{\rm d}t} = -\sum_{J \neq I} \left [ \frac{{\rm d}}{{\rm d}\bR_I} U_{IJ}(\bR_{IJ}) + \gamma_{IJ}(\bR_{IJ}) \bV_{IJ} - \sigma_{IJ}(\bR_{IJ})\xi_{IJ} \right ] \, ,
\label{eq-DPD}
\end{equation}
where $\bV_{IJ} = \bV_I - \bV_J$ is the relative velocity between CG sites $I$ and $J$, $\gamma_{IJ}$ is the DPD pair friction, and $\xi_{IJ}$ is a vector of Gaussian random variables with zero mean and unit variance.
$\gamma_{IJ}$ is often decomposed into parallel and transverse components~\cite{Junghans:2008}.
$\sigma_{IJ}$ is related to $\gamma_{IJ}$ through the fluctuation-dissipation relation: $\sigma_{IJ} = \sqrt{2\kT \gamma_{IJ}}$.
Here, we have also simplified the form of the conservative potential, $U_{IJ}(\bR_{IJ})$, to be pairwise additive and depend only on the distance between two particles, as an approximation to the MB-PMF, $U^0$. 
Of course, by further simplifying the functional form of the friction kernel to be independent of particle identity or position, we will recover a simple Langevin equation which represents the ``default'' equations of motion for bottom-up CG models in practice.

We will end this very brief introduction to the MZ formalism by pointing the interested reader to more detailed discussions of the approach: 
(i) Hij\'{o}n \etal~\cite{Hijon:2010} provide a foundational discussion of the MZ approach for parametrizing CG equations of motion;
(ii) Di Pasquale \etal~\cite{DiPasquale:2019} clearly demonstrate the link between MZ and other bottom-up coarse-graining approaches;
(iii) Izvekov~\cite{Izvekov:2013} gives a thorough account of various approximations and derivations of GLEs used within the MZ framework. 

\subsection{Friction kernel parametrization using higher-resolution simulations}
\label{ssect:friction}
Assuming a reasonable approximation, $U_{IJ}(\bR_{IJ})$, to the MB-PMF, $U^0$, the main challenge of the MZ approach is to determine an approximate friction kernel, $\Gamma$.
A conceptually simple approach to this problem is to choose a functional form for the friction kernel and then tune the parameters to reproduce local dynamical properties, e.g., velocity-velocity time correlation functions (TCFs), of the CG sites.
However, this parametrization can be practically difficult and may lead to unpredictable behavior due to uncontrolled approximations~\cite{Kauzlaric:2011}.
As a consequence, significant work has been done to develop more systematic and theoretically-founded methodologies for calculating the friction kernel.

Izvekov and Voth~\cite{Izvekov:2006} proposed to determine the frictional forces directly from simulations of a higher-resolution model using a GLE formulation for TCFs.
They approximated the lost friction due to missing degrees of freedom as the difference between the total atomistic force on each site for a given configuration and the force generated by the chosen conservative interactions for the CG model.
According to this approximation, the frictional forces were then determined from force-velocity and velocity-velocity TCFs.
Kauzaric \etal~\cite{Kauzlaric:2011} expanded upon the link between TCFs and the MZ frictional forces by proposing three distinct routes to the friction kernel based on Green-Kubo, Onsager, and Einstein-Helfand relations:
\begin{eqnarray}
\Gamma_{IJ} &=& \frac{1}{\kT} \int_0^{t} {\rm d}t' \langle \delta \bF_I(0) \delta \bF_J(t')  \rangle \label{eq:frict1} \label{Green-Kubo}\\
            &=& - \frac{M_I M_J}{\kT} \frac{{\rm d}}{{\rm d}t'} \langle \bV_I(0) \bV_J(t') \rangle \mid_{t'=t} \label{eq:frict2} \label{Onsager} \\
            &=& \frac{C^P_{IJ}(t) - C^P_{IJ}(0)}{\kT t} \,\, , \label{eq:frict3} \label{Einstein-Helfand}
\end{eqnarray}
where $\langle \rangle$ denotes an ensemble average and $C^P_{IJ}(t) = \langle \bP_I (t) \bP_J (t) \rangle$.
In Equation~\ref{eq:frict1}, $\delta \bF_I(t) = \bF_I(t) - \bF^{\rm C}_I(t)$, where $\bF_I(t)$ is the total force on CG site $I$ from the higher-resolution simulation and $\bF^{\rm C}_I(t)$ is the same quantity evaluated using the chosen CG conservative potential, $U_{IJ}(\bR_{IJ})$.
Each of these approaches assumes a time scale separation between the CG (slow) and missing (fast) degrees of freedom, requiring a careful choice of the time scale parameter, $t$.
In particular, if the friction matrix is computed from unconstrained molecular dynamics simulations, the integral in Equation~\ref{Green-Kubo} vanishes at infinite times.
This is commonly known as the ``plateau problem.'' 
More recently, Markutsya and Lamm~\cite{Markutsya:2014} proposed to calculate the friction kernel from the fluctuating forces of the reference model via the fluctuation-dissipation theorem.

With any of these approaches, the question becomes how to evaluate the corresponding TCFs.
According to the MZ formalism, the fluctuating force corresponds to dynamics in $Q$ space, which we do not have direct access to.
The simplest approach is to assume that the $Q$-space and real dynamics (i.e., those sampled by the higher-resolution reference model) are equivalent, leading to the above-mentioned plateau problem when applying the Green-Kubo relation (Equation~\ref{Green-Kubo}) to calculate the frictional forces.
Hij\'{o}n \etal~\cite{Hijon:2010ix} proposed to generate $Q$-space dynamics using simulations where the CG degrees of freedom, i.e., site positions, are constrained.
The constrained dynamics can be obtained by appropriately modifying the equations of motion for the high-resolution model~\cite{Hijon:2010ix,Davtyan:2015}.
This approach avoids the plateau problem but can introduce artifacts when insufficient time scale separation exists between the $P$ and $Q$ degrees of freedom, as is often the case for high-resolution CG models.
Izvekov~\cite{Izvekov:2013,Izvekov:2017,Izvekov:2017b} has proposed an alternative approach which, by employing a distinct projection operator and representing the friction as a product of momentum-dependent Hermite polynomials, determines the frictional forces from unconstrained molecular dynamics simulations without the onset of the plateau problem.
Furthermore, very recent work from Espa\~{n}ol and coworkers~\cite{Espanol:2019} derives corrected Green-Kubo relations that avoid the plateau problem in cases where the time scale separation is not extreme.

\subsection{Dissipative Particle Dynamics (DPD)}

As described above, if one assumes that $\Gamma$ is only dependent upon CG site positions and is pairwise additive, the full GLE is transformed into the equations of motion employed in the DPD method.
Although DPD models traditionally use soft-sphere potentials to describe mesoscale structure, conservative forces derived from, for example, structure-based CG methodologies can also be used within this framework~\cite{Espanol:2017}.
Due to its practical simplicity and efficiency, this approach has emerged as a popular framework for constructing CG models that more accurately represent the dynamical properties of a higher-resolution reference model.
The MZ-DPD approach has been applied to a range of molecular systems, especially for very coarse representations of star polymers and cluster-based mappings of fluids~\cite{Hijon:2010,Lei:2010,Yoshimoto:2013,Izvekov:2014,Li:2014,Li:2015,Li:2017}.
(We note that although considerable effort has been made to derive the DPD equations from first principles for CG sites that represent clusters of nonbonded molecules~\cite{Espanol:1997b,Flekkoy:1999,Flekkoy:2000,Ayton:2004,Eriksson:2008,Eriksson:2009b,Hadley:2010,Izvekov:2014,Han:2018}, the precise connection to the microscopic dynamics remains unclear~\cite{Espanol:2017}.
In fact, there appears to be no rigorous foundation for employing conservative forces to describe interactions between such supramolecular blobs~\cite{Bock:2007}.)
As articulated very nicely in a recent paper by Deichmann and van der Vegt~\cite{Deichmann:2018}, these studies demonstrate success of the MZ-DPD approach in cases where there is a relatively clear time scale separation between the CG ($P$) and removed ($Q$) degrees of freedom.
Here we focus on applications with higher-resolution CG representations, where this time scale separation does not clearly exist.

Izvekov and Voth~\cite{Izvekov:2006} constructed one- and two-site models of liquid methanol, using force-velocity and velocity-velocity TCFs to determine scalar, but particle-dependent, friction constants.
Conservative forces were obtained from the force-matching-based multiscale coarse-graining (MS-CG) method~\cite{Noid:2008a}.
They demonstrated that the two-site representation results in accurate molecular center of mass velocity-velocity TCFs, although the diffusion constant was only reproduced within a factor of 2.
Izvekov and Rice~\cite{Izvekov:2014} later proposed an approach to match two-body force-velocity and three-body velocity-velocity TCFs for DPD models that represent a cluster of molecules with a single DPD site.
Conservative forces were again obtained with the MS-CG method.
The frictional forces were decomposed into parallel and transverse components and were not restricted to a particular functional form.
The authors examined varying resolutions for liquid nitromethane, using a Voronoi tessellation to determine clusters of molecules that map to a single CG site.
Of particular note, while coarser representations demonstrated Markovian behavior in the TCFs that could be accurately represented with the DPD models, the one-molecule-per-site mapping resulted in significant deviations in these observables due to a lack of time scale separation.
It was also demonstrated that although the parallel frictional contributions are dominant for all representations, the transverse friction contributes significantly to both viscosity and diffusion (as expected from previous investigations~\cite{Junghans:2008dt}).
In a follow-up study from the same authors~\cite{Izvekov:2015}, the importance of the transverse frictional contributions was further established.
Li \etal~\cite{Li:2014} have also performed a detailed investigation into the effect of directional friction components in the context of coarser representations for star polymers.

Tr\'{e}ment \etal~\cite{Trement:2014} have employed the approach of Hij\'{o}n~\etal ~(see Section~\ref{ssect:friction}) to derive DPD models with two distinct representations of 5 and 10 heavy atoms per CG site for liquids of $n$-pentane and $n$-decane.
Using conservative potentials obtained from direct Boltzmann inversion, subsequently fitted to an analytic exponential form, they demonstrated qualitative agreement of both velocity-velocity TCFs and diffusion constants.
In a subsequent study, Lemarchand \etal~\cite{Lemarchand:2017} applied the same approach to {\it cis}- and {\it trans}-1,4-polybutadiene and demonstrated reasonable agreement with the underlying self-diffusions and viscosities for a range of representations from 1 to 6 monomers per CG site.
They demonstrated that dynamical accuracy increases with increasing level of coarse-graining, due to better fulfilment of the Markovianity assumption.

Deichmann \etal~\cite{Deichmann:2014} applied the Green-Kubo approach (Equation~\ref{Green-Kubo}) to determine DPD friction matrices for three different molecular liquids with both one- and two-site representations, while using conservative potentials obtained with the conditional reversible work method~\cite{Brini:2011}.
While one-site models reproduced both the overall behavior of the velocity TCFs and also the diffusive dynamics reasonably well, the two-site representations failed to accurately capture dynamical properties, supposedly due to insufficient time scale separation between the CG and removed degrees of freedom.
In a follow-up study, Deichmann and van der Vegt~\cite{Deichmann:2018} investigated liquids, polymer solutions, and polymer melts based on repeat 2,2-dimethylpropane units.
They demonstrated that the diffusion in single-component liquids and polymer solutions can be accurately modeled using the MZ-DPD formalism, despite a clear lack of time scale separation.
In this case, there is not a one-to-one correspondence between reproducing TCFs and reproducing the properties of long time diffusion.
However, discrepancies in the local free-energy barriers prevent accurate modeling of activated diffusion of small molecules in polymer melts.

Overall these investigations have demonstrated that DPD is an efficient framework for adding relatively simple dissipative forces to the CG model, based on the MZ formalism.
Recent work by Espa\~{n}ol and coworkers~\cite{Espanol:2016,Faure:2017} has extended this approach to ``entropy-based'' CG models, which incorporate the internal energy as an additional CG variable, allowing simulations of thermal transport via DPD with energy conservation~\cite{Espanol:1997}.
In general, in cases with clear time scale separation between the CG and removed degrees of freedom, the resulting DPD models can accurately represent both local dynamics (e.g., velocity TCFs) as well as long time scale properties (e.g., diffusion constants).
However, for higher-resolution CG representations, where this separation clearly does not exist, the resulting discrepancies can be rather large, depending on the precise CG mapping, functional form of the frictional forces, and the chosen conservative forces.
As a result, researchers have pursued extensions of the DPD framework, which incorporate momentum- and time-dependent frictional forces to more accurately reproduce the underlying dynamics.

\subsection{Hydrodynamic-like momentum-dependent friction}

Equation~\ref{eq-GLE} demonstrates that the friction kernel is generally dependent not only on the coordinates but also the momenta of all CG sites.
Although typically associated with movement of particles through a fluid, i.e., hydrodynamics, Soheilifard \etal~\cite{Soheilifard:2011} demonstrated that momentum-dependent frictional forces can emerge even when coarse-graining a linear model without solvent---in this case, an atomically-detailed elastic network protein model mapped to a C-$\alpha$ representation.
To distinguish from the traditional case where effective interactions arise from correlations in solvent momentum, we will refer to momentum-dependent friction that depends only on the CG degrees of freedom as ``hydrodynamic-like.''
In the realm of relatively coarse representations, Lei \etal~\cite{Lei:2010} constructed DPD models with and without hydrodynamic-like contributions to the frictional force and demonstrated the necessity of momentum dependence for reproducing the flow about clusters of Lennard-Jones particles.
They indicated that hydrodynamic-like forces will be required anytime the dissipative forces are on par with the conservative forces.
They investigated two different pairwise conservative forces, one of which reproduces the radial distribution functions by construction, and demonstrated little difference in the resulting dynamics.

Markutsya and Lamm~\cite{Markutsya:2014} proposed a method for computing hydrodynamic-like interactions directly from the distribution of random forces generated by higher-resolution simulations.
They tested their method on both pure liquid water and a single glucose molecule solvated in water, using one-site representations for each molecule and the MS-CG method to determine the conservative potentials between CG sites.
The viscosities, velocity-velocity TCFs, and self-diffusion constants were all reproduced with high accuracy relative to a CG model without dissipative forces, although there were notable structural discrepancies due to the employed conservative forces.
However, because the model was not directly compared with a DPD model without hydrodynamic-like interactions, it is unclear what precise role these forces play.
More recently, Izvekov~\cite{Izvekov:2017, Izvekov:2017b} has investigated the impact of momentum dependence on the resulting dynamical properties of a CG model by considering the approach to the Markovian limit and the regime of weak momentum dependence.
He demonstrated that for a one-site model of liquid nitromethane the momentum dependence results in a much more accurate representation of the velocity TCFs, compared with the standard MZ-DPD model~\cite{Izvekov:2014}.
Moreover, the alternative projection operator~\cite{Izvekov:2013} employed in this work may be better suited for high-resolution CG models that lack clear time scale separation, since it can deal with non-Markovian dynamics and avoids the plateau problem when using the Green-Kubo relation to determine the frictional forces from unconstrained molecular dynamics simulations.

At this point, we will briefly cover more traditional modeling of hydrodynamic interactions.
In the context of the GLE approaches described above, incorporating hydrodynamics amounts to reintroducing missing degrees of freedom (e.g., solvent) in an explicit but approximate way in order to more accurately model the movement of CG sites through an implicit medium.
Zgorski and Lyman~\cite{Zgorski:2016} employed stochastic rotational dynamics, also known as multi-particle collision dynamics, along with the implicit solvent Dry Martini model to more accurately model the thermodynamics and dynamics of a lipid bilayer.
Stochastic rotational dynamics~\cite{Malevanets:1999, Gompper:2009} is a particle-based Navier-Stokes fluid simulation technique that represents the solvent with tracer particles that conserve the hydrodynamic flow.
The authors demonstrated that, with the appropriate choice of collision rules, the approach effectively corrects for known dynamical finite-size effects while generating the quasi-two-dimensional hydrodynamics of the bilayer.
The lattice Boltzmann method~\cite{Duenweg:2007}---an alternative hydrodynamics approach which employs a lattice in lieu of tracer particles to represent the solvent---has been recently applied to study peptide self-assembly~\cite{Sterpone:2015} and protein stability in crowded environments~\cite{Gnutt:2019} using a chemically-specific CG protein model.
These studies aim to probe the role of hydrodynamic interactions in these phenomena, rather than to directly match the dynamics of some underlying reference model.
Researchers have also employed concurrent multiscale schemes that incorporate hydrodynamic interactions to model flow in liquids and at interfaces~\cite{Hu:2018,Wang:2018}.

\subsection{Friction kernels with memory}

While the Markovian assumption is typically taken for practical convenience, researchers have also investigated memory effects, i.e, time dependence, in the friction kernel, $\Gamma$.
Note that, through the fluctuation-dissipation theorem, this time dependence in $\Gamma$ results in a time-dependent random force, which in practice requires implementation of a colored-noise generator.
Cao \etal~\cite{Cao:2013} demonstrated that, for modeling an ionic solution with implicit solvent, an exponential memory function is necessary for reproducing local dynamical properties, e.g., velocity TCFs, of the ions.
Yoshimoto \etal~\cite{Yoshimoto:2013} later showed that, due to a lack of time scale separation between CG and removed degrees of freedom, a non-Markovian friction kernel is necessary for reproducing viscosity in the high-density regime for a model that represents clusters of Lennard-Jones particles with a single CG site.
Using an alternative projection operator, Izvekov~\cite{Izvekov:2013} derived an expression for the memory as a function of force-force, force-position, and force-momentum TCFs.
He showed that if the projected force is decorrelated from both the CG positions and momenta, a position- and momentum-independent memory kernel is sufficient for representing the underlying dynamics.
The work from Li \etal~\cite{Li:2015} demonstrated that in the case of limited time scale separation, which can be quantified from the difference in relaxation time scales of velocity versus force TCFs, a time-correlated random force is necessary to reproduce short-time properties that are related to how the system responds to high-frequency disturbances.
Additionally, the authors proposed two different methods for efficient implementation of a colored-noise generator.
Very recently, Jung \etal~\cite{Jung:2017, Jung:2018} have proposed a generalization of the approach of Li \etal~\cite{Li:2015}, which can be interpreted as generalized Brownian dynamics with frequency-dependent friction kernels.
In contrast to the previous approach, this method is not Galilean invariant and, thus, more applicable to the movement of CG particles in an implicit environment.
The method was tested on nanocolloid suspensions, where it was demonstrated that the incorporation of time dependence is necessary for precisely reproducing the long-range hydrodynamic interactions between two nanocolloids.
Moreover, the proposed methodology demonstrates an effective speed-up on the order of $10^4$ for this particular application.

An alternative to the standard colored-noise generators was proposed by Andersen and coworkers~\cite{Davtyan:2015,Davtyan:2016}, which uses a class of nonlinear systems interacting with special heat baths, originally introduced by Zwanzig.
Within this scheme, fictitious particles are introduced which have no impact on the static properties of the system, but which couple directly to the CG degrees of freedom.
The authors proposed the method as a simple scheme for reproducing dynamical features of an underlying model, without making a rigorous connection to the MZ framework.
The approach was tested on several systems including a one-site model for ethanol dissolved in water and one- and two-site models for liquid methanol.
The resulting models displayed excellent correspondence with the TCFs of the reference simulations.
The discrepancies which did appear in these cases could be corrected by increasing the complexity of the fictitious particles.
Lei \etal~\cite{Lei:2016} subsequently demonstrated that a fictitious particle approach can be more rigorously connected to the GLE representation through a rational function approximation of the Laplace transform of the memory kernel.
More recently, Karniadakis and coworkers~\cite{Li:2017, Yoshimoto:2017} have pursued a similar auxiliary variable approach, to simplify the implementation of non-Markovian DPD.
Although these studies consider models with coarser representations (one site for a cluster of Lennard-Jones particles or an entire star polymer), the results contain potentially useful lessons for coarse-graining at a higher-resolution.
In comparison to the previous non-Markovian DPD approach from the same group~\cite{Li:2015}, the auxiliary variable approach can reduce the computational cost by a factor of $\sim$20-30.

A different approach to simplifying the parametrization and implementation of the memory kernel was proposed by Lyubimov and Guenza~\cite{Lyubimov:2011}, who employed an approximate but analytic expression for the time-dependent friction, derived from atomistic dynamics.
This work is described further in Section~\ref{subsec-poly}.

\subsection{Variational approaches}
\label{subsec-var}

Within the MZ approaches described above, the accuracy of the resulting dynamical properties of the model is dependent on how well the dissipative forces approximate those determined from MZ theory.
The precise form of these dissipative forces is implicitly dependent on both the CG representation and the conservative forces, chosen to approximate the MB-PMF.
Fu \etal~\cite{Fu:2013} showed that the accurate representation of TCFs can be very sensitively dependent on fine features of the dissipative forces, e.g., the cut-off value employed. 
While the MZ formalism is rigorously founded, it does not provide in itself a systematic route for targeted optimization of particular properties, in contrast to variational approaches developed for determination of the conservative forces~\cite{Noid:2008a,Shell:2008il}.
In the following we describe two different variational approaches that have been proposed for constructing GLE models from higher-resolution reference simulations.

Espa\~{n}ol and Z\'{u}\~niga~\cite{Espanol:2011} formulated a generalization of the relative entropy method---a variational approach for determining the conservative forces for a CG model~\cite{Shell:2008il}.
The approach employs a Markovian approximation to the Fokker-Planck equation, which is expressed in terms of the first two Kramers-Moyal coefficients, i.e., a drift vector and a diffusion tensor.
These terms can be estimated from sets of reference simulation trajectories with distinct initial states.
In particular, the authors employed a new relative entropy functional which depends on two-time stationary joint probabilities, in contrast to the time-independent probabilities of the original relative entropy method.
They also demonstrated that the same expression can be derived starting with the path probability and following a Bayesian approach, along with the appropriate (e.g., Markovian) approximations.
The variational property of the method was demonstrated by showing that the Kramers-Moyal coefficients are recovered in the limit of a full basis set---i.e., arbitrarily complex forms to represent frictional forces.
For coarser representations which demonstrate inertial dynamics, the inverse of the diffusion tensor may not exist, yielding an ill-posed optimization problem.
However, using the Bayesian formulation, a generalized force-matching approach can be derived for obtaining both conservative and dissipative forces.
This method essentially weights individual configurations within the force-matching functional by taking into account the relative contributions from thermal noise.

Inspired by this approach, Dequidt and Canchaya~\cite{Dequidt:2015} devised a numerical scheme for parametrizing a DPD model with pairwise interactions.
Using a Bayesian formulation, the authors derived an optimization function, which can be analytically solved for models with limited complexity.
The resulting method is again a generalized force-matching (later termed as ``trajectory matching''), where a least squares fit of the average force (weighted by a frictional contribution) over a short trajectory is performed.
The method can be extended to simultaneously target the average pressure of the system, placing a restriction on the form of the conservative potential.
The approach was tested on a one-site model of n-pentane, allowing direct comparison with the work of Tr\'{e}ment \etal~\cite{Trement:2014}, who applied the Green-Kubo relation (Equation~\ref{Green-Kubo}) to determine the dissipative forces.
The static properties of the resulting model accurately represent that of the reference simulation, not only for the pure liquid but also for liquid-vapor phase equilibria, when employing reference data from multiple temperatures~\cite{Canchaya:2016}.
However, significant discrepancies appear in the dynamical properties.
In particular, the diffusion constant and TCFs demonstrate similar errors to the model of Tr\'{e}ment \etal~\cite{Trement:2014}, indicating a limitation in the functional form of the frictional forces rather than a limitation of the optimization procedure.
Similar results were demonstrated by Kempfer \etal~\cite{Kempfer:2019} applying the same methodology to a polymer melt, highlighting the difficulty of simultaneously matching both short and long time scale properties in more complex systems.
Considering these discrepancies, extensions to more flexible basis functions for the frictional forces will help to assess the capabilities of the approach.

Harmandaris and coworkers~\cite{Kalligiannaki:2015,Harmandaris:2016} have recently formulated an elegant theoretic framework which expands on the variational approaches for equilibrium coarse-graining~\cite{Noid:2008a,Shell:2008il}.
In addition to generalizing the force-matching (i.e., MS-CG) method to nonlinear mappings and recasting structure-based methods in a probabilistic framework~\cite{Kalligiannaki:2015}, the authors have proposed a variational inference approach for the non-equilibrium regime~\cite{Harmandaris:2016}.
This approach avoids the time discretization of the previous methods by employing the ``relative entropy rate''---a full path-space entropy functional for continuous-time representations~\cite{Katsoulakis:2013}.
This work provides a detailed investigation into the proposed framework, elucidating the connections between force-matching-based and relative-entropy-based non-equilibrium methods as well as equivalency conditions for a completely data-driven approach based on generic time-series data.
To the best of our knowledge, the method has not yet been tested on molecular simulation models.

\subsection{Application to proteins}
Guenza and coworkers have developed the Langevin equation for protein dynamics (LE4PD) method~\cite{Copperman:2014} by applying the projector operator technique to describe the system dynamics in terms of the $\alpha$-carbons along the protein backbone.
The resulting model represents protein fluctuation dynamics using an overdamped Langevin equation along with structural information about the important metastable configurations of the protein as input.
An extended Rouse-Zimm model is employed at the residue level, which specifically accounts for the difference in hydrodynamic interactions depending on the relative exposure to the solvent.
In contrast to the standard Rouse-Zimm model, LE4PD allows for global anisotropy and sequence specificity while incorporating internal dissipation and free-energy barriers in a physically-motivated approach.
LE4PD takes as input either short molecular dynamics simulations or experimental conformal structures and is capable of accurately representing both global and local motions of folded protein structures without any free parameters through a mode analysis approach.
Through multiple investigations considering nine different proteins~\cite{Caballero:2007,Copperman:2014,Copperman:2015}, the authors have demonstrated that LE4PD is capable of reproducing NMR relaxation time scales as well as nuclear Overhauser effects.
The LE4PD model has also recently been employed to identify universal hierarchical scaling features underlying sequence-specific protein dynamics, resembling the directed polymer in random media model~\cite{Copperman:2017}.

\section{Time rescaling relationships}

If there exists a true time scale separation between the CG and removed degrees of freedom, then the time dependence of the frictional forces will vanish.
In this case, there are two factors which determine the complexity of the frictional forces.
First, the chosen conservative force modulates the form of the dissipative forces, although this relationship is not well understood.
Additionally, the chosen projection operator within the MZ formalism determines the partitioning between the frictional and random forces~\cite{Izvekov:2013}.
Thus, with the appropriate representation, conservative interactions, and projection operator it may be possible to remove the coordinate, momentum, and time dependence from the friction kernel, resulting in a simple Langevin description for the CG dynamical evolution.
In this case, a uniform time rescaling would recover all relevant dynamical properties at the CG level of resolution.
Note however that, even in this case, the scalar friction term generally depends on the thermodynamic state point, e.g., temperature, pressure, or density.

\subsection{Polymers}
\label{subsec-poly}

Many features of polymer systems can be understood using simple theoretical models which idealize monomer-monomer and monomer-solvent interactions.
Amongst the simplest is the Rouse model~\cite{Doi:1986}, which represents segments of the polymer (i.e., one or several monomers) as CG sites connected along the chain via harmonic springs.
These segments are assumed to move independently from one another and, in the simplest case, the sites do not interact.
This model is appropriate for chain lengths shorter than the entanglement length, beyond which more sophisticated reptation models are required.
Extensions such as hydrodynamic-like interactions between segments to better model chain dynamics in dilute solution conditions can also be applied~\cite{Doi:1986}.
These models act as an important reference point from which ``real'' polymer dynamics can be compared.
For example, in the absence of entanglement, divergence from the Rouse model implies significant correlations between segments that can arise from chain end effects, chain stiffness or semiflexibility, coupling between intra- and inter-molecular forces, or the presence of strong specific interactions~\cite{Guenza:2002,Guenza:2008hc}.
Each of these effects may lead to non-trivial frictional forces from the MZ perspective.
Although homogeneous polymeric systems exhibit static and dynamic universality beyond characteristic length and time scales~\cite{Takahashi:2017}, which can be reproduced with generic bead-spring models~\cite{Kremer:1990}, chemically-specific molecular simulation models offer advantages for making direct comparisons with experimental measurements and opportunities for extensions to more complex, e.g., heterogeneous, systems.

Harmandaris, van der Vegt, Kremer, and coworkers have performed detailed investigations into the dynamical properties generated from CG models for polystyrene-based systems with a two-site per monomer representation~\cite{Harmandaris:2006xj,Harmandaris:2007rx,Harmandaris:2007oy,Harmandaris:2009oc,Fritz:2009rg,Harmandaris:2009zf,Fritz:2011,Harmandaris:2011,Harmandaris:2014}.
The conservative forces were parametrized to reproduce local structural features of a higher-resolution model.
The authors assumed a scalar time-rescaling factor, which may depend on the molecule identity as well as the thermodynamic state point, determined by matching self-diffusion coefficients to those obtained from higher-resolution simulations.
Although the rescaling factor depends on the molecular weight (i.e., chain length), it converges for relatively short chains, allowing the quantitative prediction of dynamics for system sizes intractable for the underlying reference model.
The rescaled CG dynamics show favorable comparison with dynamic structure factors from neutron spin echo spectroscopy, rotation relaxation time scales from NMR, and dielectric spectroscopy.
This simple rescaling approach is accurate on length scales of 0.5-1~nm and time scales of approximately 100~ps, allowing the description of long time diffusion as well as local chain dynamics.
By considering distinct CG representations, the authors have demonstrated that the precise rescaling factor is a sensitive function of the CG mapping, although the overall accuracy of the rescaling was similar for the different representations.
By employing an empirically-determined Arrhenius-like time-rescaling factor as a function of concentration and temperature, they have also extended the method to describe penetrant diffusion in polystyrene melts.
However, the rescaling of individual components may differ, complicating the methodology in cases without a clear time scale separation between the dynamics of distinct components.
The time rescalings are largely transferable to higher pressures, although discrepancies emerge in local side-chain dynamics.
Remarkably, the rescaling approach results in very good agreement with respect to dielectric spectroscopy experiments that probe segmental dynamics of polystyrene/oligostyrene blends which display heterogeneous dynamics~\cite{Harmandaris:2013}.
More recently, the approach has been extended to the nonequilibrium regime (e.g., polymers under sheer flow), although difficulties arise far from equilibrium due to the entropy-enthalpy compensation in CG models.
The reader is directed to~\cite{Padding:2011} and~\cite{Harmandaris:2014} for a more detailed overview of time rescaling in CG models for polymers in the equilibrium and nonequilibrium regimes, respectively.

Salerno and Grest~\cite{Salerno:2016} performed a systematic investigation of the time-rescaling dependence on resolution and chain length for relatively high-resolution representations.
They found that the time rescaling is independent of chain length (as demonstrated earlier by Harmandaris \etal~\cite{Harmandaris:2007oy}) but can depend sensitively on the CG representation.
Ohkuma and Kremer~\cite{Ohkuma:2017} demonstrated that both the stress relaxation time scale and shear rate is recovered through time rescaling for two different models of {\it cis}-polyisoprene, with and without the incorporation of a pressure correction to the conservative force.
More recently, Volgin \etal~\cite{Volgin:2018} investigated time rescalings for nanoparticle diffusion in a polyimide-based melt employing both one- and three-site per monomer representations.
Interestingly, they found that while the uniform time rescaling breaks down for intermediate time scales, taking into account the local viscosity of the polymer chain segments (whose size is comparable with the nanoparticle size) rescues the time-rescaling approach.

The time-rescaling approach suffers from several limitations.
First, there is no guarantee that a scalar time-rescaling factor exists.
For example, Agrawal \etal~\cite{Agrawal:2016} showed that time-dependent rescaling factors are necessary for accurate prediction of polymer viscoelasticity for polyurea.
Additionally, the time rescaling will generally be a function of the system identity, thermodynamic state point, and the chosen CG representation.
For example, the time-rescaling behavior in glass-forming polymers can be related to the entropy-enthalpy compensation in CG models~\cite{Song:2018}.
Lyubimov and Guenza~\cite{Lyubimov:2011,Lyubimov:2013} have addressed this issue through an analytic rescaling method which connects monomer-level dynamics with global diffusion of a polymer chain.
Using this approach, they have considered both a monomer-level representation, i.e., a bead-spring model, and a soft-sphere representation which treats the entire polymer chain with a single CG site.
The radius of gyration of the polymer as well as a hard-sphere parameter, which describes the excluded volume of monomer units, are taken as input.
Using a clever combination of MZ theory, the Ornstein-Zernike equation from liquid state theory, and a model for cooperative polymer chain dynamics, the authors were able to reconstruct the local dynamics of a polymer chain that is consistent with the overall diffusive motion of the polymer.
They demonstrated that for polyethylene and polybutadiene this approach is capable of reproducing the dynamic structure factors generated from united-atom molecular dynamics simulations.
Moreover, they have shown that the input parameters can be approximated using a freely-rotating chain model with only a small loss in accuracy as the system approaches the highly-entangled regime.

\subsection{Liquids}

There is a long history of work relating structural or thermodynamic properties with the emergent dynamical properties of liquids~\cite{Cohen:1959,Adam:1965,Rosenfeld:1977}.
For example, free volume theories relate the available space for molecular packing with dynamic coefficients for transport properties~\cite{Starr:2002,Mittal:2007}.
Perhaps most notably, Rosenfeld demonstrated that a range of transport properties (diffusivity, viscosity, thermal conductivity) are directly related to the excess entropy of the liquid~\cite{Rosenfeld:1999}.
These relationships have been validated for a wide class of liquids~\cite{Saika-Voivod:2001,Sastry:2001,Goel:2009,Krekelberg:2009,Dyre:2018} but typically only appear when the appropriate transformations to reduced thermodynamic variables are applied.

Armstrong and Ballone~\cite{Armstrong:2012} proposed that the Rosenfeld relationship could be utilized to interpret CG diffusion time scales in liquids.
They demonstrated the linear relationship between the excess entropy and the diffusion constant for a one-site model of water.
While the results demonstrate the potential to predict CG time scales, they also highlight the potential limitations of simple rescaling approaches for CG dynamics.
If distinct chemical groups have different excess entropies, then a single rescaling is not possible, and consistent dynamics cannot be easily extracted.
One predominant reasoning for the existence of thermodynamic-kinetic relationships in liquids is that all liquids can be considered to be perturbations of the venerable soft-sphere model~\cite{Likos:2001}.
Shell~\cite{Shell:2012} leveraged this perspective to optimize CG interaction potentials based on a minimization of the relative entropy with respect to a higher-resolution model.
Using this approach, he obtained the optimal soft-sphere scaling exponents, which in turn determine the time-rescaling factors for a set of binary Lennard-Jones systems.
This rescaling resulted in a collapse of self-diffusivities onto a master curve for a range of temperatures, although the relationship breaks in the supercooled regime.
We note that Accary and Teboul~\cite{Accary:2012} have shown that time rescalings for CG models of low-temperature liquids may be limited to correcting long-time diffusion at the cost of accurately representing the mechanism of local diffusion.
Velhorst \etal~\cite{Veldhorst:2014} investigated the extension of the Rosenfeld relation to larger, flexible molecules, using Lennard-Jones particles connected by harmonic springs.
They demonstrated that the harmonic bonds destroy correlations between the potential energy and the virial.
These correlations give rise to the typical structural-kinetic relationships through the generation of isomorphs on the phase diagram---curves along which structure, dynamics, and excess entropy are invariant.
In this case, isomorphs still exist, but the trend in excess entropy deviates from the other structural and dynamical properties.
The authors also make a connection to polymers, where the well known power law scaling with density indicates that isomorph theory may be applicable.

Recently, Douglas and coworkers~\cite{Xia:2018,Song:2018,Xia:2019} proposed a different method which determines temperature-dependent CG potentials by matching the Debey-Waller factor, which can be related to the relaxation time scales in the localization model of glass forming liquids.
In other words, the approach provides a calibration of the activation energy of glass formation via the cohesive interaction strength parameter in the CG potentials.
The constructed three-site CG models for {\it ortho}-terphenyl reproduced the glass transition temperature, relaxation dynamics, and self-diffusion coefficients of the underlying AA model, as well as the reduced compressibilities~\cite{Xia:2018}.
The method was also applied to one- and two-site-per-monomer models of three glass-forming polymers with distinct segmental structures, where the form of the interaction potentials were extended to be frequency dependent while taking dynamic fragility into account in order to accurately represent the viscoelastic properties of the polymers under a small-amplitude oscillatory shear force~\cite{Song:2018, Xia:2019}.
The results demonstrate very good agreement with both AA simulations and experiments, although the authors identify that differences in the CG representation (i.e., level of resolution) can complicate the comparative analysis of CG models for distinct systems.
Additionally, the temperature-dependent energy rescaling factors obtained for the three polymers, as well as for an analytically-solvable model, appear to exhibit a universal sigmoidal structure~\cite{Xia:2019}.
Overall, this work suggests that energy renormalization may provide a useful route for constructing CG models of glass-forming liquids and polymers with predictive capabilities.

\section{Free-energy-landscape (FEL) perspective}

Despite the dynamical limitations of CG models, there is extensive research which utilizes these models to gain insight into the mechanisms of structure formation in soft matter.
In particular, the energy-landscape formalism~\cite{Stillinger:1982,Stillinger:1995} provides the conceptual footing to interpret CG dynamics in a meaningful way based on metastable basins and the dominant set of barriers between these states.
Perhaps most notable is the application of energy-landscape theory for protein folding~\cite{Onuchic:1997}, which asserts that the energy landscape for globular proteins is funnel shaped, with the native state as the global minimum.
By constructing simple models based on this zeroth-order approximation, much of our foundational knowledge about the dominant driving forces of the protein folding process was obtained~\cite{Dill:1997,Clementi:2008}.
Although absolute time scales cannot be directly recovered from these models, connections between structural and dynamical properties arise through the restricted topology of the energy landscape~\cite{Leopold:1992}, resulting in increased confidence in the extraction of insight from simulated pathways.
Unlike the approaches described above, this perspective places a focus on the barrier-crossing dynamics along the dominant features of the FEL, while allowing for the diffusive motion within local free-energy minimum to differ from the true dynamics.

\subsection{Structural-kinetic-thermodynamic relationships}

\begin{figure}[htbp]
  \begin{center}
    \includegraphics[width=0.77\linewidth]{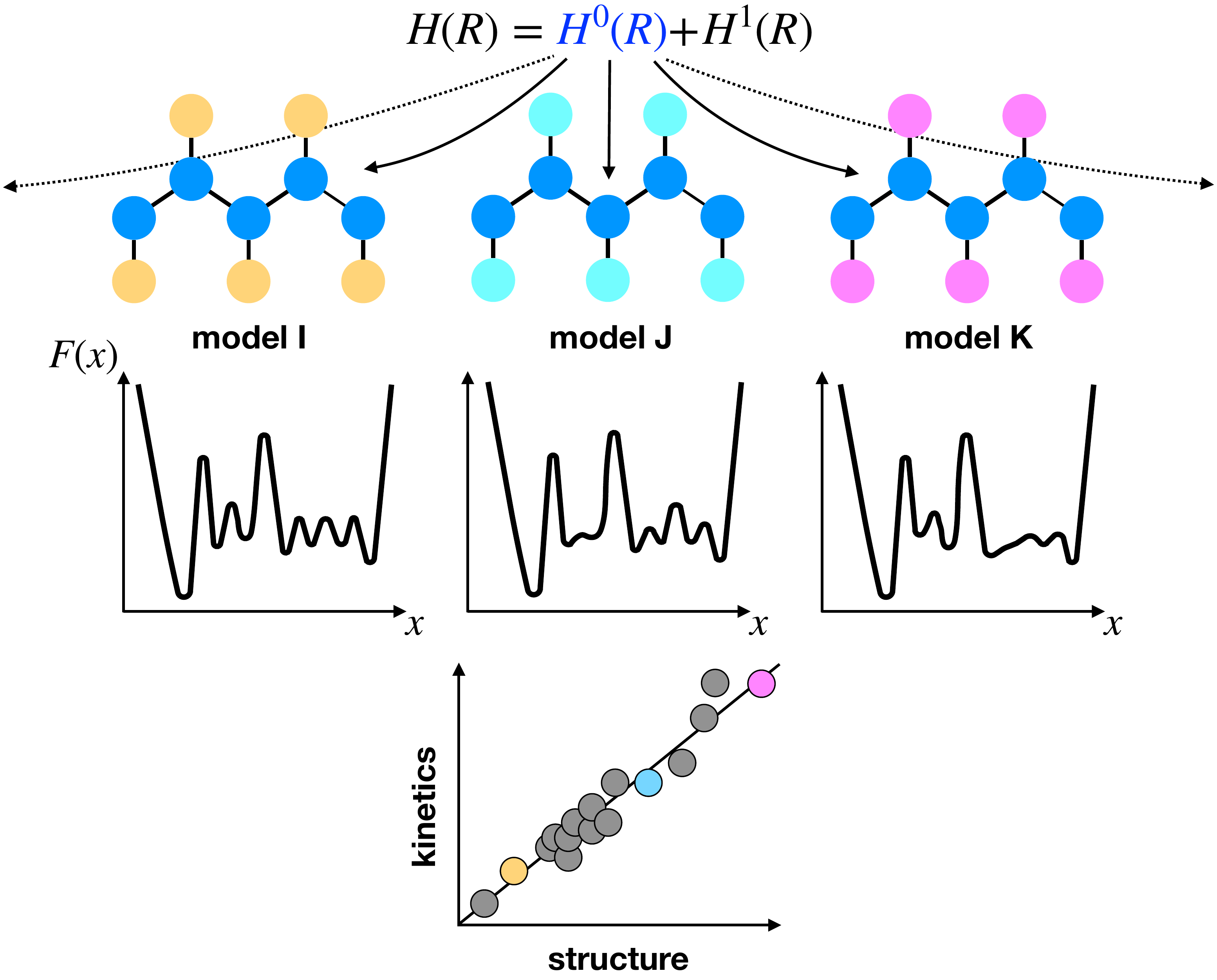}
    \caption{
Illustration of a high-throughput-inspired investigation to identify structural-kinetic relationships.
First a Hamiltonian is defined with fixed ($H^0$) and variable ($H^1$) terms.
Various models are then constructed by varying the parameters of $H^1$.
The resulting FELs are then investigated (e.g., using kinetic networks as a proxy for the features of the landscape) to identify features that are preserved due to the physical components that are present in every model (i.e., $H^0$).
Finally, structural-kinetic relationships are identified by finding correlations between various properties.
\vspace{-4mm}
}
    \label{fig:SKrelns}
  \end{center}
\end{figure}

The identification of structure-activity relationships has been widely adopted as a successful route for materials design~\cite{Bereau:2016}.
The implicit assumption of this approach is that differences in the chemical features of distinct systems give rise to dominant distinctions with respect to the FEL, which then dictate larger-scale properties of the system.
This link to the FEL qualifies the approach as a general means for understanding relationships between chemistry, structure, thermodynamics, and dynamics.
More specifically, characteristic chemical or physical properties of a system or process enforce constraints on particular features of the FEL, such that simple relationships between structural, thermodynamic, and dynamical properties may emerge~\cite{Schneider:2013,Winquist:2013,Dickson:2017,Menichetti:2017} (see Figure~\ref{fig:SKrelns}).
In the context of CG modeling, these structural-kinetic or thermodynamic-kinetic relationships are somewhat similar in principle to the time-scaling relationships described above, as they provide a handle for quantitatively interpreting the dynamical properties generated by the CG model.
Thus, such relationships might be useful for determining effective scaling relationships for building kinetic consistency into CG models, without the use of dissipative forces.

Inspired by high-throughput studies, Rudzinski and Bereau~\cite{Rudzinski:2018} have recently performed a detailed investigation into structural-kinetic-thermodynamic relationships for helix-coil transitions.
They employed a simple CG model which represents hydrogen-bonding and hydrophobic side chain interactions with simple attractive potentials, while maintaining an atomically-detailed description of excluded volume to avoid sampling of sterically-forbidden conformations.
The authors demonstrated that steric effects alone enforce a rather strict restriction of the attainable FELs, characterized through a kinetic network description.
This restriction gives rise to very simple relationships between structure and kinetics, e.g., between the average fraction of helical segments and the ratio of nucleation to elongation rates, irrespective of the specific details of the model.
The results further demonstrate a link between cooperativity (i.e., thermodynamic properties) and the topology of the kinetic network (e.g., randomness of pathways from the helix to the coil state) at a single temperature.
In a follow-up study~\cite{Rudzinski:2018b}, the same authors investigated the role of conformational entropy in these structural-kinetic relationships and found that a shift in the relationships occurs depending on the representation of steric interactions.

\subsection{Markov state models (MSMs)}

MSMs are discrete-time and -space master equation~\cite{Klippenstein:2014} models that can be parametrized directly from molecular simulation trajectories~\cite{Bowman:2014}.
The problem of constructing an MSM is akin to finding a finite-dimensional, linear approximation to the dynamical propagator of the probability density of the system, which is intimately related to the Koopman operator approach from dynamical systems~\cite{Klus:2018}.
MSMs describe the dynamics of the system as memoryless jumps between {\it microstates} which represent volume elements of configuration space (i.e., sets of entire configurations of the system), in contrast to the particle-based CG models discussed above which explicitly represent the details of individual molecules.
The microstate definition is crucial for the accuracy of the model, and is routinely determined in a semi-automated fashion using a combination of dimensionality reduction~\cite{Mu:2005,Naritomi:2011,Schwantes:2013,Perez-Hernandez:2013,Schwantes:2015,Chen:2018b,Wehmeyer:2018} and clustering~\cite{Sittel:2016,Lemke:2016,Husic:2018b} methodologies.
Given the configuration space discretization, the transition probabilities between microstates (i.e., the MSM parameters) can be inferred from the number of jumps between pairs of microstates observed in the simulation trajectory within a chosen lag time~\cite{Noe:2008,Prinz:2011b,Trendelkamp:2013}.
No{\'e} and coworkers have introduced the variational approach to conformational dynamics~\cite{Noe:2013,Nuske:2014} which generalizes this procedure to arbitrary basis function representations---beyond the indicator functions used for traditional MSMs with discrete-state representations.
More specifically, the variational scheme aims to directly approximate the eigenfunctions of the exact dynamical propagator~\cite{Klus:2018}.
A large number of other developments including cross-validation techniques~\cite{McGibbon:2015} and the incorporation of machine learning algorithms into the kinetic modeling workflow~\cite{Mardt:2018} have contributed to the rapid emergence of these kinetic models as powerful tools for extracting insight from molecular simulations, particularly popular within the atomistic simulation community.
(The reader is directed to a recent review by Husic and Pande~\cite{Husic:2018} for a detailed account of MSM methodologies and applications.)
Importantly, MSMs provide a link between microscopic details of a system and the hierarchy of long time scale processes.
Although MSMs are CG models in their own right (representing an explicit coarse-graining in both space and time), the following considers the use of these models as a tool for assisting in the parametrization or interpretation of particle-based CG models.

Rudzinski and Bereau~\cite{Rudzinski:2016b} investigated the refinement of two distinct CG peptide models to better reproduce the kinetic properties of given reference models, by altering the conservative force field to better represent the dominant barrier-crossing dynamics.
Their approach relies on an MSM-based Bayesian dynamical reweighting technique~\cite{Rudzinski:2016} which determines the minimal adjustments to the kinetic model necessary for consistency with given kinetic reference data.
By comparing this ``biased'' MSM with the MSM which most accurately represents the simulation data alone, they demonstrated that deficiencies of the CG model can be identified and subsequently targeted in a reparametrization.
For example, for a C-$\alpha$-based CG model of tetraalanine, the authors demonstrated that by allowing relatively small errors in the ensemble-averaged structural properties---the target property for the original parametrization---the refined model nearly quantitatively reproduces the ratio of mean first passage times between metastable states while also more accurately representing intermediate structures (i.e., the mechanism of transition between metastable states).
These results demonstrate that the inaccurate representation of intermediate states---in this case, the sampling of sterically-forbidden states---destroys the connection between reproducing structural properties and reproducing the hierarchy of kinetic processes.
Furthermore, this inaccuracy can be linked to the limited representation of cross correlations between degrees of freedom governing terms in the CG potential, due to the additivity of these terms in standard molecular mechanics Hamiltonians.

While higher-order interactions can be employed to reproduce these correlations in principle, parametrization and implementation of these interactions are often prohibitively expensive.
Bereau and Rudzinski~\cite{Bereau:2018} have recently proposed a scheme for reproducing complex cross-correlations, without expanding the standard molecular mechanics basis set.
Analogous to surface-hopping schemes for electronic transitions, the approach employs different CG force fields for distinct intramolecular conformations of a molecule and transitions between these force fields on the fly during the simulation.
Remarkably, by applying the scheme to the C-$\alpha$ representation of tetraalanine described above, the authors demonstrated that accurate modeling of cross correlations leads to very high accuracy in the sampling of intermediate structures between metastable states (i.e., a better description of free-energy barriers), which in turn leads to the reproduction of barrier-crossing dynamics without adjusting the intra-basin friction (i.e., without the incorporation of dissipative forces).

Chen \etal~\cite{Chen:2018} have employed MSMs in conjunction with a framework for refining simulation models by integrating information from multiple experimental measurements in order to refine a CG model to more faithfully represent the folding mechanism of the FiP35 protein.
To validate the proposed procedure, the investigation employed AA reference simulations as a proxy for experimental data.
Through the consideration of one- and two-site per amino acid representations, the results highlight the importance of local frustration in reproducing the dominant features of the folding landscape.
Very recently N\"{u}ske \etal~\cite{Nuske:2019} have proposed a ``spectral matching'' method, which attempts to parametrize a CG potential to accurately represent the long time scale processes by targeting the dominant eigenfunctions of the system's dynamical propagator.
This method utilizes the variational approach to conformational dynamics~\cite{Noe:2013,Nuske:2014} along with methods for parameter estimation for stochastic dynamics~\cite{Crommelin:2011}.
The method was demonstrated on a set of toy models as well as a CG model of alanine dipeptide which represents the system along the $\phi/\psi$ dihedral angles.
The results demonstrate that the method is capable of refining a force-matching-based model to more accurately reproduce the characteristic time scales of the underlying system.
The proposed methodology appears to provide both a rigorous framework for the FEL perspective of CG dynamics presented in this section and also a practical alternative to the variational approaches described in Section~\ref{subsec-var}.

\section{Outstanding challenges through representative examples}

There are countless investigations using CG models which assess either quantitative dynamical properties or the qualitative pathways of structure formation generated by a particular model. 
Here, we discuss a few representative topics or studies which, in our opinion, highlight important aspects of the outstanding challenges in the context of interpreting and correcting CG dynamical properties.

As we have discussed in many examples above, coarse-graining of molecular liquids leads to disparate dynamical properties, including both short time scale TCFs and long time scale diffusion properties.
Correcting both scales simultaneously can be challenging for both MZ and time-rescaling approaches, depending on the chemical features of the system.
These difficulties are especially severe for ionic liquids, where strong non-covalent interactions lead to both structural and dynamic heterogeneity that is notoriously difficult to represent~\cite{Jeong:2011}.
The presence of two species alone negates the use of traditional time-rescaling approaches, due to inhomogeneous rescaling for the different components~\cite{Karimi-Varzaneh:2010}.
Moreover, the simultaneous reproduction of local packing through the delicate balance of conservative forces while also correcting for lost friction presents a formidable challenge for equilibrium coarse-graining approaches and MZ-based methodologies.
While certain dynamical properties may be universally conserved~\cite{Pal:2017}, the consistent representation of transport properties, essential for connection with real-world applications, remains difficult even for highly optimized models that accurately reproduce both structural and thermodynamic properties~\cite{Moradzadeh:2018}.

Mukherjee \etal~\cite{Mukherjee:2017} have illustrated a related but distinct example of challenging time rescaling for a CG model of azobenzene-based liquid crystals.
The CG model was developed by applying traditional structure-based techniques to a reference simulation in the supercooled state.
As a result, the model not only accurately reproduces structural properties of the amorphous phase, but also very closely matches the transition temperature from the amorphous to smectic phase with respect to the underlying AA model~\cite{Mukherjee:2013}.
Both the AA and CG models display two mechanisms of translocation between layers in the smectic phase: (i) the straight mechanism, where a molecule jumps straight between two layers without reorientation, and (ii) the parking-lot mechanism, where a molecule rotates 90~deg when exiting a layer and remains in the interlayer spacing for some period of time before entering the second layer.
Despite the accurate ensemble-averaged properties, the CG model not only exhibits distinct time rescalings for various diffusion processes (each of the two translocation pathways, interlayer diffusion in the smectic phase, and diffusion in the amorphous phase) but also displays qualitatively incorrect propensities of the two translocation mechanisms.
These two examples demonstrate fundamental challenges that will appear for modeling a wide range of industrially-relevant systems including heterogeneous polymeric systems~\cite{Zeng:2008} (e.g., polymer nanocomposites) as well as for investigating the pathways of molecular assembly processes, which may evolve along very rugged free-energy landscapes with predominant kinetic traps~\cite{Sun:2017}.

One might naively assume that the problem of interpreting CG dynamics would be necessarily more complicated for the conformational dynamics of macromolecules in solution compared with the diffusion processes of liquids.
However, there are cases where chemically-specific CG models for biomolecules are capable of generating rates and pathways of structure formation that are consistent with experimental measurements.
For example, the OxDNA model~\cite{Sulc:2012}---a two-site per nucleotide model for DNA parametrized to reproduce the melting temperatures of short duplexes---generates relative rates of duplex hybridization~\cite{Ouldridge:2013,Schreck:2015}, strand displacement~\cite{Srinivas:2013}, and hairpin formation~\cite{Mosayebi:2014} that appear directly comparable to experiments.
Apparently, the melting-curve-based parametrization probes the dominant energetic driving forces (i.e., base pairing interactions) for a whole range of larger-scale structure formation. 
On the other hand, despite the success of simple CG protein models in clarifying the essential physics of the protein folding process, the difficulties of interpreting CG dynamical properties plague the characterization of sequence-specific folding, even with rather specialized CG models.
For example, Habibi {\it{et al.}}~\cite{Habibi:2016} investigated the folding processes generated by three different CG models: a C-$\alpha$ G\={o} model~\cite{Clementi:2008}, an AA G\={o} model~\cite{Whitford:2009}, and the AWSEM model~\cite{Davtyan:2012}.
They demonstrated that the three different models provide disparate descriptions of the forced unfolding process of a 110 residue peptide, despite the capability of all models to fold the peptide to the proper native structure.
Furthermore, the fidelity of the folding process, with respect to an AA reference simulation, did not correlate with the complexity of the model.

\section{Discussion and outlook}

The last five to ten years have seen a noticeable shift towards developing dynamically-consistent CG models, including (i) the development of advanced Mori-Zwanzig-based methodologies for applications without clear time scale separation, (ii) the repurposing of well-known thermodynamic-kinetic relationships for time rescaling of diffusion processes in liquids, (iii) the identification of structural-kinetic-thermodynamic relationships for interpreting the conformational dynamics of macromolecules, and (iv) the utilization of kinetic models to characterize and correct the hierarchy of long time scale kinetic processes.
(See Figure~3 for a schematic overview of these approaches.)

\begin{figure}[htbp]
  \begin{center}
    \includegraphics[width=0.8\linewidth]{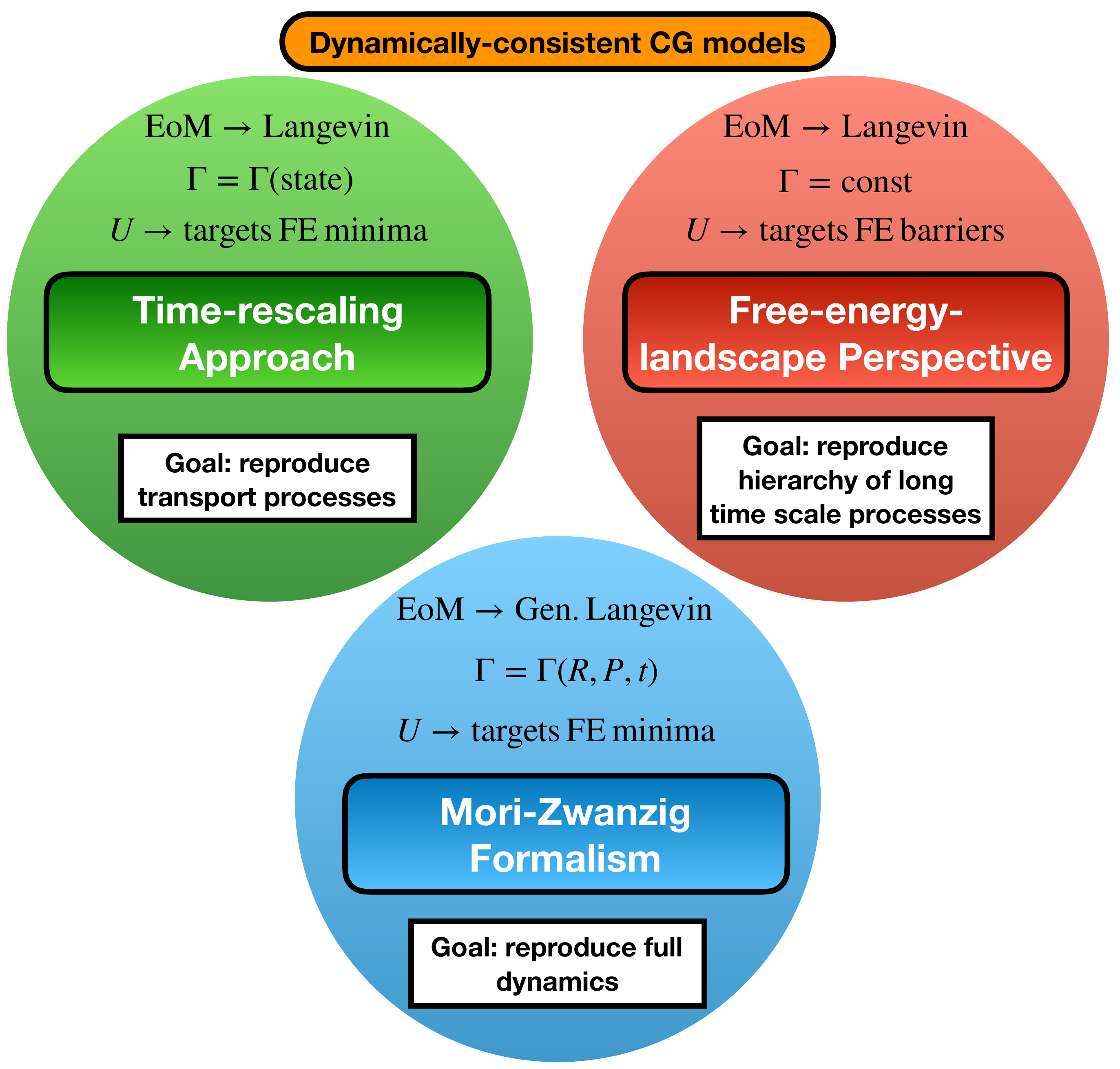}
    \caption{
Schematic of three distinct approaches to CG dynamics.
(Top left) The time-rescaling approach attempts to reproduce long time sale target properties (e.g., the diffusion constant) of the underlying system.
A simple Langevin equation motion (EoM) is applied, without the incorporation of dissipative forces.
However, an effective scalar friction constant $\Gamma$, which can depend on the identity and thermodynamic state of the system, is employed a posteriori in order to calibrate the dynamics against the target property.
The conservative potential is usually obtained via some standard coarse-graining approach, which aims to reproduce ensemble-averaged
 properties of the system (i.e., effectively targeting free-energy minima).
(Top right) The free-energy-landscape perspective attempts to reproduce the dominant hierarchy of long time scale processes of the underlying system.
A simple Langevin EoM is applied, without the incorporation of dissipative forces, which corresponds to the incorporation of a scalar friction coefficient $\Gamma$.
The conservative potential is determined in order to not only reproduce features of the free-energy minima, but to also accurately represent the dominant barriers along the free-energy landscape.
(Bottom) The Mori-Zwanzig approach attempts to reproduce the full dynamics of the underlying system by employing a generalized Langevin EoM to model the movement of CG sites.
The corresponding friction kernel $\Gamma$ generally depends on the coordinates and momenta of all CG sites, as well as on time.
The conservative potential is usually obtained via some standard coarse-graining approach, which aims to reproduce ensemble-averaged properties of the system (i.e., effectively targeting free-energy minima).
\vspace{-4mm}
}
    \label{fig:Summary}
  \end{center}
\end{figure}

While the Mori-Zwanzig formalism has been traditionally applied to describe the motion of a system along one or a small set of coarse observables, it has now been established as a practical tool for extending the development of CG models beyond the parametrization of conservative forces.
At the same time, there is still much to be learned about the appropriate complexity of configuration- and momentum-dependent frictional forces needed to accurately reproduce the dynamical properties for particular systems and CG representations.
Additionally, systematic methods are required for choosing the corresponding set of hyperparameters (e.g., cut-off distances), which can dramatically affect the resulting short- and long-time dynamical behavior~\cite{Fu:2013}.
In the case that there is no clear time scale separation between the CG and removed degrees of freedom, time-dependent random forces are necessary to reproduce detailed features of the time correlation functions, although simpler models may be adequate for reproducing longer time scale properties~\cite{Deichmann:2018}.
Efficient parametrization and implementation of colored-noise thermostats, such as the use of fictitious beads coupled to the CG degrees of freedom~\cite{Davtyan:2015}, are of critical importance for alleviating the computational cost of these models.
The development of variational approaches that target dynamical properties of the model represents a significant step towards not only more systematic investigations into the optimization of these models but also a better fundamental understanding of the required forces for reproducing particular properties.
However, despite the development of several elegant theoretical frameworks~\cite{Espanol:2011,Harmandaris:2016,Nuske:2019}, there seems to be a lack of applications to molecular systems.
It is currently unclear whether this is due to practical difficulties of the approaches or simply a lack of adoption from external researchers.
One limiting factor in successful applications of these approaches may be that there is a lack of methodology for systematically choosing the CG mapping that maximizes the time scale separation between the CG and removed degrees of freedom.

In addition to the optimization of the dissipative forces, a major goal moving forward should be to gain a better understanding of the interplay between the chosen CG representation and the conservative and dissipative forces required for reproducing particular dynamical properties.
Although some studies~\cite{Lei:2010} have shown that the conservative forces have little impact on resulting dynamical properties, others~\cite{Deichmann:2018} indicate that the conservative and dissipative forces are more strongly coupled as the complexity of the system increases, e.g., through the introduction of distinct components.
The CG mapping determines the idealized free-energy surface (i.e., the many-body potential of mean force) for representing the structural and thermodynamic properties of the underlying system.
However, the set of chosen interactions between CG sites typically provide a rather poor approximation to this complex landscape (Figure~1).
In practice, one often attempts to represent the dominant features of this landscape by targeting certain lower-dimensional properties (e.g., pair correlation functions).
The incorporation of dissipative forces is similar in the sense that, given the many-body potential of mean force as the CG conservative potential, there exists a perfect representation of the CG dynamics (i.e., the full generalized Langevin equation), although in practice we can only attempt to describe the dominant features of the ``true'' friction kernel.
However, the conditional dependence on the CG conservative force complicates the construction of the appropriate set of dissipative forces.
In other words, similar to the dependence of the optimal conservative force on the CG representation (i.e., mapping), the optimal dissipative forces depend both on the conservative force and the CG representation.
There have been shockingly few investigations into the optimization of the CG mapping or conservative potential to reduce the complexity of the dissipative forces~\cite{Guttenberg:2013}.
Therefore, future work should focus on the continued development of methods which couple the choice of the CG representation with the optimization of the conservative and dissipative forces.

Perhaps it is useful to look towards the free-energy-landscape perspective for inspiration, where researchers have devised heuristic methods for modeling the dominant kinetic features, dictated by the major barriers along the landscape.
In comparison to the rigorous bottom-up methodologies, these approaches take advantage of large length and time scale behavior and appear to provide a greater range of transferability, not unlike the difference between bottom-up and top-down methods that target equilibrium properties~\cite{Noid:2013uq}.
The use of kinetic models (e.g., Markov state models) for characterizing the hierarchy of long time scale processes and associated pathways of structure formation in a tractable fashion seems to be an advantageous route forward.
Additionally, the identification of structural-kinetic-thermodynamic relationships can significantly simplify the interpretation of the dynamical properties generated by CG models.
While some relationships determine universal time-rescaling factors for a whole class of systems, others may be limited to a particular type of process or within a limited range of molecular identities.
Nevertheless, these connections can provide confidence in the interpretation of properties away from the system or state point of parametrization, enhancing the predictive capabilities of CG models.
Furthermore, these relationships (or lack thereof) inform the ``representability'' of the model, i.e., the capability of the model to simultaneously reproduce various structural, thermodynamic, and dynamical properties.
In particular, the connection between reproducing various properties may be compromised by the limitations of the standard molecular mechanics functions employed to represent interactions between CG sites.
The adoption of extended interaction sets or alternative models~\cite{Bereau:2018} that can more accurately represent cross correlations between CG degrees of freedom will generally simplify the problem of CG dynamics, regardless of the approach taken.

At the end of the day, bottom-up CG models rely on a higher-resolution reference model, which precludes predictive capabilities \emph{unless} the variation of model parameters over system identity and thermodynamic state point is known a priori.
The combination of bottom-up formalisms with analytic theories~\cite{Lyubimov:2011} as well as data-driven techniques~\cite{Lei:2016} can assist in predicting or learning, respectively, these variations, propelling the utility of CG models. 
Overall, interfacing between the approaches discussed in this review (Figure~3) should lead to improved methods for interpreting and correcting the dynamical properties generated by CG simulation models, providing us not only with quantitatively interpretable models but also with confidence in the fidelity of predicted pathways.


\vspace{6pt} 


\subsection*{Acknowledgements}
JFR thanks Tristan Bereau, Will Noid, and David Rosenberger for critical review of the manuscript.
JFR is especially grateful to Tristan Bereau for many thoughtful discussions and guidance. 

\bibliography{references_PSU,references_MPIP}

\end{document}